# Spectral analysis of ultraluminous X-ray pulsars with models of X-ray pulsars

Manish Kumar,★ Rahul Sharma ⓘ, Biswajit Paul and Vikram Rana

*Raman Research Institute, C.V. Raman Avenue, Bangalore 560080, Karnataka, India*

## ABSTRACT

A fraction of the ultraluminous X-ray (ULX) sources are known to be accreting neutron stars as they show coherent X-ray pulsations with pulse periods ranging from ∼1–30 s. While initially thought to host intermediate-mass black holes, ULXs have since been recognized as a diverse class of objects, including ULX pulsars. These pulsars require models specifically tailored to account for their unique accretion physics, distinct from those used for Galactic black hole binaries. The X-ray spectra of all Galactic accreting X-ray pulsars (including sources in the Magellanic Clouds) are dominated by a high-energy cut-off power law and some of the sources show a soft excess, some emission lines, cyclotron absorption features, etc. In this work, we undertake a comprehensive analysis of the broad-band X-ray spectra of five ULX pulsars using simultaneous *XMM–Newton* and *NuSTAR* observations and show that their X-ray spectra can be effectively described by spectral models, similar to those used for the local accretion-powered X-ray pulsars. A soft excess is detected in all the sources which is also consistent with the local X-ray pulsars that have low absorption column density. We have marginal detection or low upper limit on the presence of the iron K-alpha emission line from these sources, which is a key difference of the ULX pulsars with the local accreting X-ray pulsars. We discuss the implication of this on the nature of the binary companion and the accretion mechanism in the ULX pulsars.

## 1 INTRODUCTION

Highly luminous off-centre X-ray sources exceeding the Eddington luminosity for stellar mass compact stars were detected with the Einstein Observatory in several nearby galaxies (Fabbiano 1989). Subsequent investigations with the *ROSAT*, *XMM–Newton*, and *Chandra* X-Ray Observatory confirmed that these high-luminosity X-ray sources are distinct from the active galactic nuclei typically associated with galaxies. This discovery marked the identification of a new class of X-ray sources, which came to be known as ultraluminous X-ray (ULX) sources (Makishima et al. 2000; Fabrika et al. 2021; King, Lasota & Middleton 2023). Currently, about 1800 ULX sources are known (Walton et al. 2021).

The luminosity of ULXs is often in the range of $10^{39}$–$10^{41}$ erg s$^{-1}$ in the X-ray energy band, which is much greater than Eddington luminosity for stellar mass compact objects. Such extraordinary luminosities are explained by assuming that the compact objects in ULXs are either intermediate-mass black holes (IMBHs) with masses ranging from $10^3$ to $10^5$ M$_\odot$, accreting at sub-Eddington rates, or stellar-mass black holes (with masses between 5 and 80 M$_\odot$), accreting at super-Eddington rates (Pintore et al. 2017; Fabrika et al. 2021).

However, subsequent timing analyses of these ULX sources have unveiled pulsations in some of them, suggesting that the compact object is a neutron star (∼1.4 M$_\odot$) in those sources. These pulsating ULX sources are referred to as ultraluminous X-ray pulsars (ULXPs). The first ULX source in which pulsations were discovered was M82 X-2 (Bachetti et al. 2014). Subsequently, coherent pulsations were identified in several other sources, including NGC 7793 P13 (Israel et al. 2017b), NGC 5907 ULX1 (Israel et al. 2017a), M51 ULX-7 (Castillo et al. 2020), NGC 300 ULX1 (Carpano et al. 2018), and NGC 1313 X-2 (Sathyaprakash et al. 2019). Apart from these extragalactic sources, three ultraluminous X-ray pulsars were also identified in Small Magellanic Cloud (SMC) and in our own Galaxy, namely SMC X-3 (Tsygankov et al. 2017; Koliopanos & Vasilopoulos 2018), RX J0209.6−7427 (Chandra et al. 2020; Vasilopoulos et al. 2020), and Swift J0243.6+6124 (Wilson-Hodge et al. 2018; Beri et al. 2021; Bykov et al. 2022).

Phenomenological modelling of the X-ray spectra of ULXs shows them to have two spectral components: a power law with a photon index of 1.5 that has an exponential cut-off below 10 keV, and a soft excess that dominates below 1 keV (Kaaret, Feng & Roberts 2017). Similar to the galactic black hole binaries, the relative strength of the soft and the hard components vary from source to source and in different spectral states of the same source. ULXPs tend to exhibit continuum-dominated spectra with a few line features. Blue shifted absorption lines have been detected in a few sources indicating outflow (Pinto, Middleton & Fabian 2016; Kosec et al. 2017). Rest-frame emission lines have been detected in rare cases (Pinto et al. 2017), with the exception of the galactic ULXs (Tsygankov et al.

★ E-mail: manishk@rrimail.rri.res.in

2017; Vasilopoulos et al. 2020). Absence of strong iron K-alpha emission line is a notable difference between the extragalactic ULXs and the known Galactic accreting compact stars. Most Galactic high-mass X-ray binaries show a large absorption column density of hydrogen, in excess of $10^{22}$ cm$^{-2}$, which is not seen in the ULXs. With a few exceptions, owing to the limited photon statistics, multiple phenomenological spectral models can often be fitted to the same spectrum of ULXs.

In previous studies, ULXPs spectra have typically been explored using models designed for various types of accreting mechanisms onto the compact object. Among these, certain models have become widely used as phenomenological approaches to explain the spectra of ULXPs (Fürst et al. 2017; Walton et al. 2018a; Brightman et al. 2022). We tabulated the detailed properties of ULXPs spectra studied previously in Table A1 of Appendix A (Fürst et al. 2017; Carpano et al. 2018; Walton et al. 2018a; Robba et al. 2021; Brightman et al. 2022). The softer part of the spectrum is usually considered a thermal emission, modelled using either an accretion disc blackbody (diskbb) or sometimes a combination of disc blackbody (diskbb) and accretion disc blackbody with a power-law radial temperature profile (diskpbb). For diskpbb, the inner disc temperature usually falls within $kT_{in} \sim 1.5$–$3.0$ keV with a power-law index $p \sim 0.55$–$1.3$ of the radial temperature profile, referred to as the hotter component. This hotter component dominates above the energy 1.0 keV and is usually associated with the emission from the innermost region of the supercritical accretion disc (Robba et al. 2021). For the low-temperature disc blackbody model (diskbb), the inner disc temperature is typically lower, around $kT_{in} \sim 0.17$–$0.45$ keV, and referred to as a cooler component. This cooler component dominates below 1.0 keV and is often associated with the emission from a larger part of the accretion disc outside the spherization radius. The cooler component likely represents the emission from the outflowing winds from the disc due to the mass-loss caused by radiation pressure (Robba et al. 2021; Brightman et al. 2022). The harder part of the spectrum is often fitted with either a simple power-law model or a power law with an exponential cut-off, which accounts for the emission from the accretion columns onto the neutron star. The cut-off energy typically lies around $E_{cutoff} \sim 5$–$8$ keV. In ULXPs spectra, the column density is usually found to be $N_H \sim 3$–$85 \times 10^{20}$ cm$^{-2}$, which is generally associated with the absorption from the local surrounding medium or the galactic medium (Carpano et al. 2018; Robba et al. 2021; Brightman et al. 2022).

In the accretion-powered pulsars, the strong magnetic field of the neutron star interrupts the disc at the magnetospheric radius (Pringle & Rees 1972; Davidson & Ostriker 1973). Even at the Eddington accretion rate, the magnetospheric radius is quite large and the disc temperature due to viscous dissipation is not high enough to emit in the X-rays. The spectral model, a multicolour disc blackbody used to fit the soft excess of these ULXPs, is not applicable for accretion onto a high-magnetic field neutron star. In a few galactic X-ray pulsars (e.g. Her X-1, 4U 1626–67; Hickox, Narayan & Kallman 2004), and some pulsars in the Magellanic Clouds (SMC X–1, LMC X–4; Paul et al. 2002; Naik & Paul 2004; Sharma et al. 2023a) which have relatively low absorption column density, the soft excess is explained as reprocessed emission from the inner accretion disc, that is heated by the X-rays from the neutron star. A single-temperature blackbody component fits the soft excess in these sources quite well.

In contrast to previous studies that primarily employed models mentioned in the above paragraphs when analysing the X-ray spectra of ULXPs (e.g. Fürst et al. 2017; Walton et al. 2018a; Brightman et al. 2022), we aim to examine the broadband spectra of the ULXPs with models that are most commonly used to describe the spectra of the Galactic accretion-powered X-ray pulsars (including sources in the Magellanic Clouds). To achieve this, we employ a single spectral model comprising of a power law with a high-energy exponential cut-off and a blackbody component for the soft excess that describes the spectra of most local accreting pulsars (e.g. White, Swank & Holt 1983; Paul et al. 2002). The emission model is modified by an absorption component characterizing the absorption of X-rays due to the material enveloping the compact object and along our line of sight. The key focus is on the blackbody temperature ($kT_{BB}$), which as observed in most local X-ray pulsars, tends to exhibit low values indicative of a soft excess in the X-ray spectrum (Paul et al. 2002; Hickox et al. 2004; Naik & Paul 2004; Sharma et al. 2023a).

By applying this unified spectral model to analyse the observed spectra of ULXPs, we aim to draw direct comparisons with local X-ray pulsars. Additionally, we seek to determine the upper limits of equivalent widths for iron K-alpha line in the ULXPs and examine their potential correlations with absorption column density and compare them with local pulsars. This comparison allows us to shed light on the unique characteristics of ULXPs while exploring their similarities with the local pulsars and trying to probe the nature of their companion stars.

## 2 OBSERVATIONS AND SOURCE SELECTION

For this work, we have selected the brightest (quasi-)simultaneous observations from *XMM–Newton* (Jansen et al. 2001) and Nuclear Spectroscopic Telescope Array (*NuSTAR*: Harrison et al. 2013) of ULXPs. Our analysis focuses on ULXP sources, namely NGC 7793 P13, NGC 5907 ULX1, M51 ULX-7, NGC 300 ULX-1, and NGC 1313 X-2, utilizing the available simultaneous X-ray data from both observatories (detailed properties in Table 1). We did not include M82 X-2 in this spectral study because of contamination from M82 X-1 source within close proximity. The ULXPs do not show significant variation in spectra over time (e.g. Fürst et al. 2017). Therefore, we considered only observations which are simultaneously or quasi-simultaneously observed by both instruments with sufficiently good statistics. Details of observation are given in Table 2.

### 2.1 NGC 7793 P13

NGC 7793 P13 is an X-ray binary object situated in a nearby galaxy NGC 7793 ($D = 3.9$ Mpc; Karachentsev et al. 2004). The pulsations in NGC 7793 P13 were discovered by Israel et al. (2017b) in the X-ray emission using *XMM–Newton* data with a spin period of 0.42 s and spin-up was also observed with a period derivative of $\dot{P} \sim -4.0 \times 10^{-11}$ s s$^{-1}$ using two one-year apart observations. The maximum observed luminosity of this source is $L \sim 1.6 \times 10^{40}$ erg s$^{-1}$ in the energy range of 0.3–10 keV (Israel et al. 2017b). This source is also observed in the optical and UV bands. By studying the optical counterparts of this source, Motch et al. (2014) concluded that the companion is B9Ia spectral type star having a mass of 18–23 M$_\odot$ and orbital period is about 64 d, observed with radial velocity of He II emission (Motch et al. 2014).

### 2.2 NGC 5907 ULX1

NGC 5907 ULX-1 is located in edge-on spiral galaxy NGC 5907 at a distance ($D = 17.06$ Mpc; Tully et al. 2016). The pulsations in NGC 5907 ULX1 were discovered by Israel et al. (2017a) in observations of epoch 2003 (*XMM–Newton*) and 2014 (*XMM–Newton* and *NuSTAR*) with spin period of 1.427 and 1.137 s, respectively.

**Table 1.** Observed properties of ULXPs.

| Sources | Distance (Mpc) | $P_{\rm spin}$ | $P_{\rm orb}$ | $P_{\rm sup-orb}$ | $L_{\rm peak}$ (erg s$^{-1}$) | References |
| --- | --- | --- | --- | --- | --- | --- |
| M82 X-2 | 3.6 | 1.37 s | 2.5 d | – | $4.9 \times 10^{39}$ | Bachetti et al. (2014) |
| NGC 7793 P13 | 3.9 | 0.42 s | 64 d | – | $\sim 10^{40}$ | Karachentsev et al. (2004), Israel et al. (2017b), Fürst et al. (2016) |
| NGC 5907 ULX1 | 17.06 | 1.13 s | 5 d | 78 d | $\sim 10^{41}$ | Tully, Courtois & Sorce (2016), Israel et al. (2017a), Fürst et al. (2017) |
| M51 ULX-7 | 8.6 | 2.8 s | 2 d | 38 d | $\sim 10^{39-40}$ | McQuinn et al. (2016), Castillo et al. (2020), Brightman et al. (2022) |
| NGC 300 ULX1 | 1.88 | 31.6 s | – | – | $4.7 \times 10^{39}$ | Gieren et al. (2005), Carpano et al. (2018) |
| NGC 1313 X-2 | 4.2 | 1.5 s | – | – | $9 \times 10^{39}$ | Tully et al. (2016), Sathyaprakash et al. (2019), Robba et al. (2021) |

A spin-up trend was also observed with a period derivative of $\dot{P} \sim -9.6 \times 10^{-9}$ s s$^{-1}$ and $\dot{P} \sim -5.0 \times 10^{-9}$ s s$^{-1}$ in years 2003 and 2014, respectively. The observed peak luminosity of this source is $L \sim (2.2 \pm 0.3) \times 10^{41}$ erg s$^{-1}$. The neutron star orbits around its companion star in a nearly circular orbit with an orbital period of 5.3 d and a projected semimajor axis of 2.5 light-seconds (Israel et al. 2017a). A superorbital period of 78.1 d was also observed in this source (Walton et al. 2016).

### 2.3 M51 ULX-7

M51 ULX-7, located in spiral galaxy M51 ($D = 8.6$ Mpc; McQuinn et al. 2016), is another ULX source in which pulsations are observed with a spin period of 2.8 s and spinning-up with a period derivative of $\dot{P} \sim -1.5 \times 10^{-10}$ s s$^{-1}$. The observed luminosity of this source is $L \sim (6$–$8) \times 10^{39}$ erg s$^{-1}$ in the energy ranges of 0.3–10 keV. In this X-ray binary, the neutron star is orbiting around its companion star (an OB-type giant/supergiant star having mass greater than 8 M$_\odot$) with an orbital period of about 2 d (Castillo et al. 2020). A superorbital period of $\sim 38$ d was also observed in this source (Brightman et al. 2022).

### 2.4 NGC 300 ULX1

NGC 300 ULX1 (earlier known as SN 2010da) is located in the spiral galaxy NGC 300 at a distance of 1.88 Mpc (Gieren et al. 2005). Pulsations are observed in NGC 300 ULX1 using data of *XMM–Newton* and *NuSTAR* during epoch 2016 with a spin period of 31.6 s and spinning-up with an unusually high period derivative of $\dot{P} \sim -5.56 \times 10^{-7}$ s s$^{-1}$. The observed luminosity of this source is $L \sim 4.7 \times 10^{39}$ erg s$^{-1}$ in the energy ranges of 0.3–30 keV (Carpano et al. 2018).

### 2.5 NGC 1313 X-2

NGC 1313 X-2 is located in barred spiral galaxy NGC 1313 at a distance 4.2 Mpc (Tully et al. 2016). The pulsations in this source were observed at a period of $\sim 1.5$ s with the *XMM–Newton* observation during the epoch of 2017 (Sathyaprakash et al. 2019). These observed pulsations were weak compared to other ULXP sources (Sathyaprakash et al. 2019). The observed maximum luminosity of this source is $L \sim 9 \times 10^{39}$ erg s$^{-1}$ in the energy ranges of 0.3–10 keV (Robba et al. 2021).

## 3 DATA REDUCTION

### 3.1 *XMM–Newton*

*XMM–Newton* is a space X-ray observatory of the European Space Agency. There are three scientific instruments aboard *XMM–Newton* (Jansen et al. 2001): European Photon Imaging Camera (EPIC), Reflection Grating Spectrometer, and an Optical Monitor. There are three EPIC cameras: two MOS-CCD cameras and one PN-CCD camera and the energy range for EPIC is 0.3–10 keV (Strüder et al. 2001; Turner et al. 2001). In this work, we used only EPIC-pn data for our spectral analysis.

For the reduction of *XMM–Newton* data, we used *XMM–Newton* Science Analysis System (SAS) version 20.0.0 and followed SAS data analysis threads. First, the raw events files were extracted using `epproc` for the listed observations. Then, we performed the background flaring correction depending on the observations, if required. For removing background flare, we extracted PN light curve in the energy ranges of 10–12 keV and removed the time intervals in which count rate is greater than 0.45 for NGC 7793 P13, 0.6 for M51 ULX-7, 0.4 for NGC 1313 X-2, and 0.35 for both observations of NGC 300 ULX1. The source events were extracted from the circular region of radius, centred at the source coordinates, mentioned in Table 2. The background events were extracted from a region twice the size of the source area, positioned away from the source location. Then, we grouped the spectrum by selecting minimum counts of 20 and oversample equal to 3 using task `specgroup`. Instrument response file (rmf) and ancillary response file (arf) are generated using the tasks `rmfgen` and `arfgen`, respectively.

### 3.2 *NuSTAR*

The *NuSTAR* observatory is the first focusing high-energy X-ray telescope launched by NASA. It comprises two identical focal plane modules, FPMA and FPMB, and operates in the energy range of 3–78 keV (Harrison et al. 2013). We used both FPMA and FPMB data for our spectral analysis. For the reduction of *NuSTAR* data, we used standard NUSTARDAS software and latest calibration files (version: 20220118). The clean event files were generated using NUPIPELINE (v: 0.4.9) and then NUPRODUCTS was used for extracting the barycentre-corrected light curves and spectrum files for both instruments FPMA and FPMB. The source events were extracted from a circular region of radius mentioned in Table 2 centred at the source position and background events were selected from the area doubled by source area region away from the source. Then, spectral files were grouped to a minimum S/N ratio including background using FTGROUPPHA with groupscale between 3.0 and 5.0 depending upon the statistics of data.

## 4 RESULTS

To perform a comprehensive spectral analysis, we have considered an energy range spanning from 0.3 to 20.0 keV, employing XSPEC software for spectral fitting (Arnaud 1996). We simultaneously fitted the EPIC-pn, *NuSTAR*-FPMA, and *NuSTAR*-FPMB spectra. A `constant` was included to account for the cross-calibration between these three instruments and fixed it to 1 for FPMA, and allowed it to vary for others. Our spectral model consists of an absorbed power law with an exponential high-energy cut-off (highecut) with a blackbody component added, to investigate the soft excess in the spectrum. A similar model is used to study Galactic X-ray binary pulsars (e.g. White et al. 1983; Paul et al. 2002; Doroshenko et al. 2017; Sharma et al. 2023a). For the blackbody component, we use bbodyrad model which gives information about the radius of the emission region from where soft excess X-ray emission originates. To model interstellar absorption, tbabs was used, with the abundances set to *wilm* (Wilms, Allen & McCray 2000) and cross-sections set to *vern* (Verner et al. 1996). Additionally, a Gaussian component was included to examine the equivalent width of the iron line. The final model used in XSPEC is cons*tbabs*(bbodyrad+highecut*powerlaw+ gaussian). In Fig. 1, we present the energy spectra fitted with the model combination presented above for all the five selected sources. The source name and observation ID are marked in each figure. The best-fitting spectral parameters are given in Table 3.

Table 2. The X-ray observation log of ULXPs used in this work.

| Sources | Co-ordinates | Instrument | Date | Obs. ID | Good time exposure (ks) | Source region | Flaring correction | References |
|---|---|---|---|---|---|---|---|---|
| NGC 7793 P13 | RA = $23^h57^m50.9^s$ Dec. = $-32°37'26.6''$ | *NuSTAR* *XMM–Newton* | 20-05-2016 20-05-2016 | 80201010002 0781800101 | 106 22 | 50'' 30'' | – Yes | Walton et al. (2018a), Israel et al. (2017b) |
| NGC 5907 ULX1 | RA = $15^h15^m58.62^s$ Dec. = $56°18'10.3''$ | *NuSTAR* *NuSTAR* *XMM–Newton* | 09-07-2014 12-07-2014 09-07-2014 | 80001042002 80001042004 0729561301 | 57 56 38 | 40'' 35'' 35'' | – – No need | Fürst et al. (2017), Israel et al. (2017a) |
| M51 ULX-7 | RA = $13^h30^m1.02^s$ Dec. = $47°13'43.8''$ | *NuSTAR* *XMM–Newton* | 10-07-2019 11-07-2019 | 60501023002 0852030101 | 169 59 | 30'' 20'' | – Yes | Brightman et al. (2022), Castillo et al. (2020) |
| NGC 300 ULX1 | RA = $00^h55^m4.85^s$ Dec. = $-37°41'43.5''$ | *NuSTAR* *XMM–Newton* *XMM–Newton* | 16-12-2016 17-12-2016 19-12-2016 | 30202035002 0791010101 0791010301 | 163 84 40 | 50'' 30'' 30'' | – Yes Yes | Carpano et al. (2018) |
| NGC 1313 X-2 | RA = $3^h18^m22.00^s$ Dec. = $-66°36'4.3''$ | *NuSTAR* *XMM–Newton* | 16-12-2012 16-12-2012 | 30002035002 0693850501 | 100 91 | 40'' 30'' | – Yes | Robba et al. (2021) |

For NGC 7793 P13, the spectral analysis includes one *NuSTAR* and one *XMM–Newton* simultaneous observation from 2016. The 0.3–20 keV spectra can be well fitted with absorbed power law with exponential cut-off ($\chi^2$/dof ∼ 562.9/551). Addition of a soft component (bbodyrad) further improved fit statistics ($\chi^2$/dof ∼ 549.7/549) and gave a blackbody temperature of $kT_{BB} = 0.17^{+0.12}_{-0.04}$ keV with normalization of norm$_{BB} = 16.8^{+81.4}_{-13.4}$, cut-off at energy $E_C = 5.0^{+0.3}_{-0.8}$ keV and folding energy at $E_F = 5.6^{+0.2}_{-1.4}$ keV with power-law index $\Gamma = 0.99^{+0.06}_{-0.32}$.

NGC 5907 ULX1 has been observed at two distinct epochs, occurring in 2013 and 2014, during which it was observed simultaneously or nearly simultaneously by both *XMM–Newton* and *NuSTAR* instruments. The results obtained from examining the spectra in 2013 and 2014 are in agreement with the absence of spectral variability (Fürst et al. 2017). Therefore in this report, we only present the analysis of observations in epoch 2014 which have higher statistics. In 2014, we had nearly simultaneous *NuSTAR* observations separated by ∼3 d. We carefully checked both observations and found no significant variation between them. Hence, we simultaneously fitted both *NuSTAR* observations with the *XMM–Newton* observation. The spectra can be well fitted with an absorbed power law with an exponential cut-off without a blackbody component. Amongst the five sources, NGC 5907 ULX1 has the largest absorption column density and perhaps a low temperature blackbody component is therefore not directly detectable. If a blackbody component is added to the model, it is not possible to constrain both its temperature and normalization. Therefore, we fitted the spectrum with the blackbody temperature fixed at 0.27 keV, which is the best-fitting value of the initial unfrozen fit and found the upper limit on blackbody normalization to be 2.2. Also, we obtained cut-off at energy $E_C = 5.1^{+0.5}_{-0.8}$ keV and folding energy $E_F = 8.1 \pm 0.9$ keV with power-law index of $\Gamma = 1.46^{+0.04}_{-0.15}$.

For M51 ULX-7, simultaneous observations from *NuSTAR* and *XMM–Newton* during 2019 were utilized. The *NuSTAR*-FPMA spectrum was used in the energy ranges of 3–20 keV. However, *NuSTAR*-FPMB spectrum was found to be dominated by background above 15 keV, so we restricted FPMB spectrum in the 3–15 keV energy range. The spectral fit with the absorbed blackbody and power law with an exponential cut-off model yielded cut-off at energy $E_C = 6.0^{+0.8}_{-0.7}$ keV and folding energy $E_F = 6.7^{+1.6}_{-1.3}$ keV with power-law index of $\Gamma = 1.39^{+0.12}_{-0.13}$. The soft excess blackbody component is found to be at temperature $kT_{BB} = 0.217^{+0.025}_{-0.026}$ keV with norm$_{BB} = 6.1^{+4.7}_{-2.2}$.

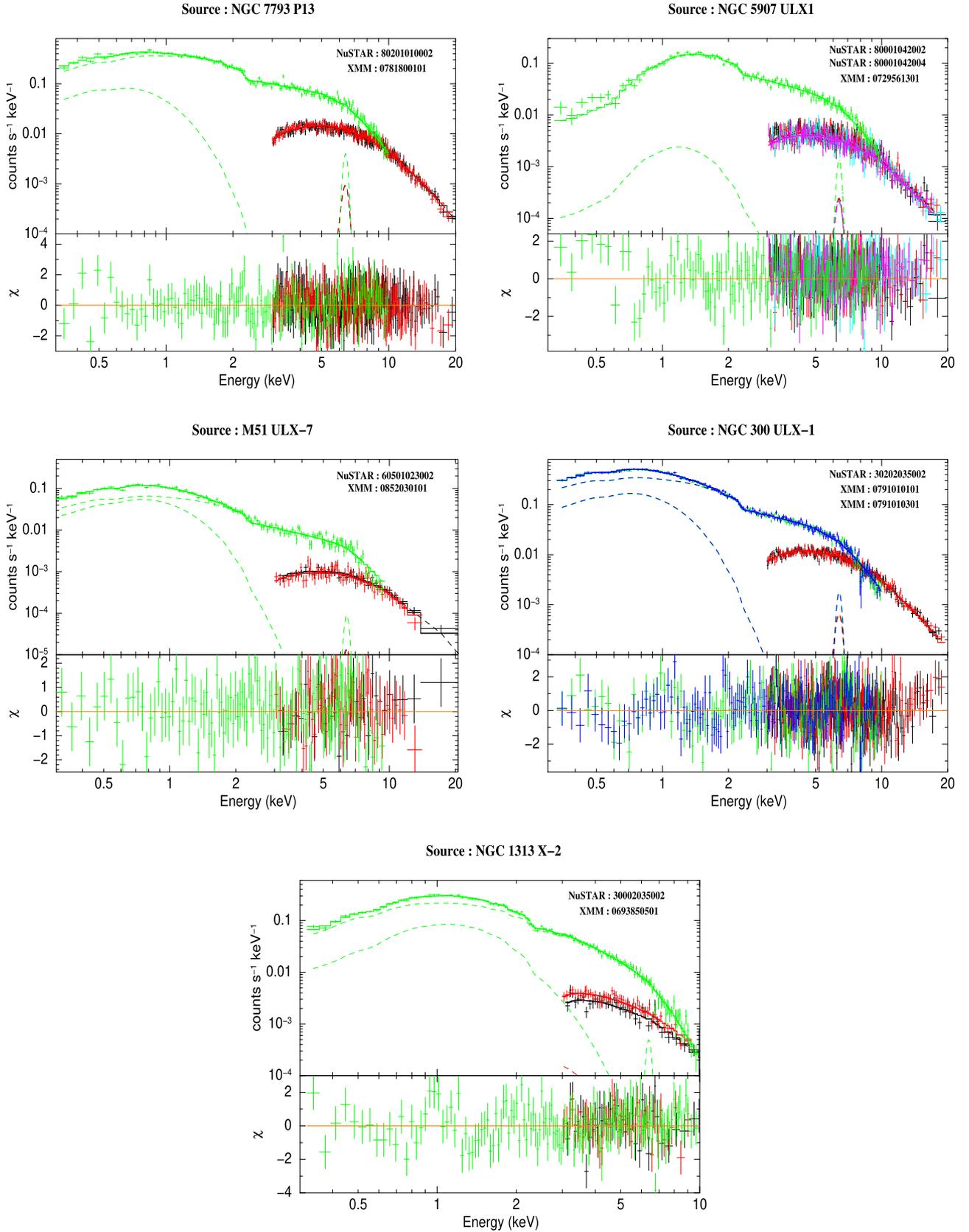

**Figure 1.** Best-fitting spectrum of ULXPs with simultaneous or quasi-simultaneous observations from *XMM–Newton* and *NuSTAR*. Green colour data points correspond to the EPIC-pn spectra. Black and red data points represent the spectra from *NuSTAR* FPMA and FPMB, respectively. For NGC 5907 ULX1, the second *NuSTAR* observation is presented in cyan and magenta colours for FPMA and FPMB, respectively. For NGC 300 ULX-1, the second *XMM–Newton* observation is presented in blue colour.

Table 3. Best-fitting spectral parameters for the ULXPs sources using model `cons*tbabs*(bbodyrad+highecut*powerlaw+gaussian)`.

| Component | Parameter | Units | NGC 7793 P13 | NGC 5907 ULX1 | M51 ULX7 | NGC 300 ULX1 | NGC 1313 X2 |
|---|---|---|---|---|---|---|---|
| TBabs | $N_H$ | $10^{20}$ cm$^{-2}$ | $9.1^{+4.1}_{-5.1}$ | $72.5^{+4.5}_{-3.3}$ | $9.9^{+2.9}_{-2.6}$ | $6.6^{+0.9}_{-1.0}$ | $21.8^{+5.1}_{-6.1}$ |
| BBodyrad | $kT_{BB}$ | keV | $0.17^{+0.12}_{-0.04}$ | $0.27^a$ | $0.217^{+0.025}_{-0.026}$ | $0.211^{+0.013}_{-0.012}$ | $0.36^{+0.05}_{-0.03}$ |
| BBodyrad | norm$_{BB}$[b] | | $16.8^{+81.4}_{-13.4}$ | < 2.2 | $6.1^{+4.7}_{-2.2}$ | $26.1^{+6.9}_{-5.0}$ | $2.1^{+1.5}_{-1.4}$ |
| BBodyrad | $R_{BB}$ | $10^8$ cm | $1.6^{+2.3}_{-0.9}$ | < 2.5 | $2.1^{+0.7}_{-0.4}$ | $1.0^{+0.1}_{-0.1}$ | $0.6^{+0.8}_{-0.4}$ |
| highecut | $E_{cutoff}$ | keV | $5.0^{+0.3}_{-0.8}$ | $5.1^{+0.5}_{-0.8}$ | $6.0^{+0.8}_{-0.7}$ | $4.7^{+0.3}_{-0.3}$ | $3.2^{+0.2}_{-0.2}$ |
| highecut | $E_{Fold}$ | keV | $5.6^{+0.2}_{-1.4}$ | $8.1^{+0.9}_{-0.9}$ | $6.7^{+1.6}_{-1.3}$ | $6.3^{+0.3}_{-0.3}$ | $3.0^{+0.5}_{-0.5}$ |
| powerlaw | $\Gamma$ | | $0.99^{+0.06}_{-0.32}$ | $1.46^{+0.04}_{-0.15}$ | $1.39^{+0.12}_{-0.13}$ | $1.29^{+0.05}_{-0.07}$ | $1.34^{+0.25}_{-0.40}$ |
| powerlaw | norm | $10^{-4}$ | $4.53^{+0.41}_{-2.33}$ | $3.56^{+0.39}_{-0.62}$ | $0.90^{+0.19}_{-0.16}$ | $6.27^{+0.40}_{-0.47}$ | $3.78^{+1.32}_{-1.35}$ |
| gaussian | norm[c] | $10^{-7}$ | $23^{+18}_{-16}$ | $7^{+9}_{-7}$ | $0.6^{+5.1}_{-0.6}$ | $16^{+9}_{-9}$ | $3^{+5}_{-3}$ |
| EQW(upper limit) | – | eV | 74.9 | 80.4 | 88.5 | 60.7 | 78.6 |
| BB Lum.[d] | $L_{BB}$ | $10^{39}$ erg s$^{-1}$ | $0.28^{+0.16}_{-0.12}$ | <3.11 | $1.18^{+0.11}_{-0.19}$ | $0.22^{+0.02}_{-0.02}$ | $0.78^{+0.14}_{-0.14}$ |
| Total Lum.[d] | $L_X$ | $10^{39}$ erg s$^{-1}$ | $13.6^{+0.3}_{-0.3}$ | $118^{+6}_{-6}$ | $9.5^{+0.7}_{-0.6}$ | $3.07^{+0.06}_{-0.07}$ | $6.1^{+0.5}_{-0.4}$ |
| Constant | $C_{FPMA1}$ | | 1.00 | 1.00 | 1.00 | 1.00 | 1.00 |
| | $C_{FPMB1}$ | | $1.05^{+0.03}_{-0.03}$ | $1.04^{+0.07}_{-0.06}$ | $0.96^{+0.08}_{-0.08}$ | $1.01^{+0.02}_{-0.02}$ | $1.15^{+0.09}_{-0.08}$ |
| | $C_{FPMA2}$ | | – | $0.97^{+0.07}_{-0.06}$ | – | – | – |
| | $C_{FPMB2}$ | | – | $1.01^{+0.07}_{-0.06}$ | – | – | – |
| | $C_{PN1}$ | | $0.87^{+0.03}_{-0.02}$ | $0.93^{+0.05}_{-0.05}$ | $0.89^{+0.07}_{-0.06}$ | $0.81^{+0.02}_{-0.02}$ | $0.99^{+0.07}_{-0.06}$ |
| | $C_{PN2}$ | | – | – | – | $0.98^{+0.02}_{-0.02}$ | – |
| $\chi^2$/dof | | | 549.7/549 | 517/573 | 198/206 | 796.6/656 | 229.3/233 |

[a]Blackbody temperature is fixed to values at 0.27 keV.
[b]norm$_{BB}$ = $(R_{km}/D_{10\,kpc})^2$.
[c]Line centre and width were fixed to values of 6.4 and 0.10 keV, respectively.
[d]Unabsorbed X-ray luminosity in the energy range of 0.3–20 keV.

For NGC 300 ULX-1, we analysed one *NuSTAR* and two *XMM–Newton* simultaneous observations during epoch 2016. We obtained cut-off at energy $E_C = 4.7 \pm 0.3$ keV and folding energy $E_F = 6.3 \pm 0.3$ keV with power-law index $\Gamma = 1.29^{+0.05}_{-0.07}$. A soft excess blackbody component was found at temperature $kT_{BB} = 0.211 \pm 0.012$ keV with normalization of norm$_{BB} = 26.1^{+6.9}_{-5.0}$.

Lastly, for NGC 1313 X-2, the spectral analysis focuses on one *NuSTAR* and one *XMM–Newton* simultaneous observation from 2012 in the energy range of 0.3–10 keV. Above 10 keV, both *NuSTAR*-FPMA and FPMB spectra were dominated by the background. The best-fitting model gave cut-off at energy $E_C = 3.2 \pm 0.2$ keV and folding energy $E_F = 3.0 \pm 0.5$ keV with power-law index $\Gamma = 1.34^{+0.25}_{-0.40}$. The soft excess component was found at temperature $kT_{BB} = 0.36^{+0.05}_{-0.03}$ keV with normalization of norm$_{BB} = 2.1^{+1.5}_{-1.4}$.

In this work, we also tried to study any correlation between the equivalent width of iron K-alpha line and column density following a similar study for galactic X-ray binaries done by Pradhan, Bozzo & Paul (2018). Although there is no clear detection of the iron line but to find the 90 per cent upper limit on equivalent width, we used a Gaussian component with a fixed line centre at 6.4 keV and line width at 0.1 keV. After fitting, the value of the upper limit (90 per cent) on equivalent width was found to be 74.9, 80.4, 88.5, 60.7, and 78.6 eV for sources NGC 7793 P13, NGC 5907 ULX1, M51 ULX-7, NGC 300 ULX-1, and NGC 1313 X-2, respectively. The column density for these sources was found to be $9.1^{+4.1}_{-5.1}$, $72.5^{+4.5}_{-3.3}$, $9.9^{+2.9}_{-2.6}$, $6.6^{+0.9}_{-1.0}$, and $21.8^{+5.1}_{-6.1}$ in units of $10^{20}$ atoms cm$^{-2}$, respectively (see Table 3). Similar low absorption column density is also observed in previous studies as mentioned in Table A1.

## 5 DISCUSSION

Since the discovery, the ULXs have been believed to be accreting IMBHs or stellar mass black holes with super Eddington accretion. But now a fraction of them are known to host neutron stars accreting at super-Eddington rates. Even larger number of ULXs may have neutron stars as the compact star, pulsation from which may not have been detected for various reasons (faintness, decoherence of a fast pulsar due to orbital motion, etc.; King & Lasota 2016; King et al. 2023). In this work, we carried out a broad-band spectral analysis of five ULXPs using (near)simultaneous observations from *XMM–Newton* and *NuSTAR* from the observations with the highest statistical quality of data for each source. The resulting broad-band spectra help us evaluate the suitability of the pulsar spectral model for fitting the spectra of ULXPs. We fitted all the spectra with an absorbed power law with a high-energy exponential cut-off to account for emission from the accretion column and a blackbody component for the soft excess. This model has been used extensively to fit the X-ray spectra of local accreting pulsars.

The spectra of all five sources were well fitted with this model. The spectral analysis revealed common characteristics among them. They exhibited a photon index in the range of ~1–1.5 with an exponential cut-off at ~3–6 keV. A similar cut-off was also observed in previous studies for sources NGC 5907 ULX1, NGC 300 ULX1, and M51 ULX-7 as mentioned in Table A1 (Fürst et al. 2017; Carpano et al. 2018; Brightman et al. 2022). Most local X-ray pulsars show a cut-off at about 10 keV, however, a similar spectral cut-off at ~5 keV was observed from SMC X-1 (Paul et al. 2002). ULXP sources showed low absorption column density (in the range of 0.6–7 × $10^{21}$ cm$^{-2}$) and a soft excess at ~0.2–0.3 keV except for NGC 5907 ULX1. Similar soft excess at ~0.1–0.3 keV has been observed in local X-ray pulsars like SMC X-1 (Paul et al. 2002), LMC X-4 (Paul et al. 2002; Naik & Paul 2004; Sharma et al. 2023a), Her X-1 (dal Fiume et al. 1998) and 4U 1626–67 (Beri, Paul & Dewangan 2015), with low absorption column density.

In previous studies, the soft excess observed in ULXPs has often been modelled as emission from the outer regions of the accretion disc, beyond the spherization radius, using XSPEC model diskbb (Robba et al. 2021; Brightman et al. 2022). In our analysis, however, we used a blackbody component (bbodyrad) to model the observed soft excess, assuming it as the reprocessed emission of hard X-rays originating near the neutron star surface by an optically thick material, most likely located near the inner edge of the accretion disc. We have measured the blackbody radius, which can be approximated as the radius of the reprocessing shell from which the soft energy photons are re-emitted, to be about $R_{BB} \sim 10^8$ cm for all five sources. We also calculated the co-rotation radius ($R_\Omega$) for all five sources as mentioned in Table 4 using equation (1),

$$R_\Omega = \left(\frac{GMP_{spin}^2}{4\pi^2}\right)^{1/3} = 1.5 \times 10^8 \, P_{spin}^{2/3} \, M_{NS}^{1/3} \text{ cm}, \quad (1)$$

where $P_{spin}$ is the spin period of ULXP. For all five sources, the inner disc radius ($R_{BB}$) is smaller than or comparable to the co-rotation radius ($R_\Omega$).

**Table 4.** Corotation radius ($R_\Omega$), magnetic field ($B$), and blackbody temperature ($kT_{BB}$), estimation for ULXPs using equation (1), (2), and (3), respectively. We utilize the luminosity ($L_X$), blackbody radius ($R_{BB}$), and spin period ($P_{spin}$) values mentioned in Table 3 and Table 1, respectively.

| Source | $R_{BB}(10^8 \text{cm})$ | $B(10^{13}\text{G})$ | $kT_{BB}$ (keV) | $R_\Omega(10^8\text{cm})$ |
|---|---|---|---|---|
| NGC 7793 P13 | 1.6 | 3.80 | 0.45 | 0.94 |
| NGC 5907 ULX1 | < 2.5 | 24.41 | 0.62 | 1.82 |
| M51 ULX-7 | 2.1 | 5.11 | 0.36 | 3.33 |
| NGC 300 ULX1 | 1.0 | 0.79 | 0.39 | 16.77 |
| NGC 1313 X-2 | 0.6 | 0.46 | 0.60 | 2.20 |

Assuming that the radius of the blackbody component is same as the magnetospheric radius ($R_m$) such that $R_{BB} \sim R_m$, we can estimate the magnetic field strength of the neutron star, using the relation from Frank, King & Raine (2002),

$$R_m \sim 0.5 \, R_A \sim 1.5 \times 10^8 \times M_{NS}^{1/7} R_6^{10/7} L_{37}^{-2/7} B_{12}^{4/7} \text{ cm}, \quad (2)$$

where $R_A$ is a standard Alfven radius, $M_{NS}$ is neutron star mass in units of solar mass $M_\odot$, $R_6$ is NS radius in units of $10^6$ cm, $L_{37}$ is the X-ray luminosity in units of $10^{37}$ erg s$^{-1}$, and $B_{12}$ is the surface magnetic field of neutron star in units of $10^{12}$ G. Assuming neutron star mass $M_{NS} \sim 1.4 \, M_\odot$ and radius $R_{NS} \sim 10$ km, we found that the surface magnetic field of the ULXPs should be in the range of $B \sim 10^{13}$–$10^{14}$ G for the inner disc radius not to be larger than the co-rotation radius.

A local X-ray pulsar, SMC X-1, exhibits a peak luminosity of $1.2 \times 10^{39}$ erg s$^{-1}$ during its peak superorbital intensity phase (Pradhan, Maitra & Paul 2020), which is significantly above the Eddington limit for a neutron star accretor. Remarkably, the observed luminosities of four out of five ULXPs (with the exception of NGC 5907 ULX1) fall within a factor of 2.5–11 of SMC X-1's peak luminosity. Other local X-ray pulsars, such as SMC X-3 (Tsygankov et al. 2017) and Swift J0243.6+6124 (Beri et al. 2021), have also shown super-Eddington luminosities during outbursts, indicating similar super-Eddington accretion onto neutron stars. Moreover, the surface magnetic field of SMC X-1 has been estimated to be $B \sim 4.2 \times 10^{12}$ G (Pradhan et al. 2020), which is comparable to the magnetic field estimated for ULXPs like NGC 1313 X-2, and lies within a factor of 2–12 for three other ULXPs as given in Table 4. These findings suggest that the accretion processes in ULXPs may be similar to that observed in local X-ray pulsars. However, the precise magnetic field strengths of ULXPs still remain uncertain.

For instance, Chen, Wang & Tong (2021) studied the magnetic field of ULXPs using various accretion torque models. They calculated magnetic fields of ULXPs of the order $B \sim 10^{13}$–$10^{14}$ G, using accretion torques models from Ghosh & Lamb (1979), Wang (1995); Kluźniak & Rappaport (2007), and Campbell (2012), and $B \sim 10^{12}$ G using accretion torques model from Lovelace, Romanova & Bisnovatyi-Kogan (1995). Additionally, Meng, Pan & Li (2022) estimated magnetic fields of $B \sim 10^{13}$–$10^{14}$ G for ULXPs (excluding NGC 7793 P13) using the Ghosh & Lamb (1979) accretion torque model, with NGC 7793 P13 exhibiting a magnetic field of B $\sim 10^{12}$ G. Similarly, for M51 ULX-7, Hu, Ueda & Enoto (2021) estimated a magnetic field of $B \sim 10^{13}$ G. Indeed, there are studies where the estimated magnetic field strengths for ULXPs are lower than what we estimated. For instance, Middleton et al. (2019) studied the spectrum of M51 ULX-8, where a Cyclotron Resonant Scattering Feature was observed, and they derived an upper limit on the dipole magnetic field of $10^{12}$ G. Additionally, King & Lasota (2019) concluded that magnetar field strengths are not required to describe ULXPs. These studies suggest that there may be variability in magnetic field strengths among ULXPs, and the observed phenomena may not always require extremely strong magnetic fields.

Using the value of $L_X$ and $R_{BB}$ mentioned in Table 3, we estimated the temperature of the blackbody emission as mentioned in Table 4 using the following equation (3),

$$\frac{L_X}{4\pi R_{BB}^2} = \sigma T^4, \quad (3)$$

where $\sigma$ is the Stefan–Boltzmann constant. For all five sources, the blackbody temperature is calculated to be 0.36–0.62 keV, which is close to measured values as mentioned in Table 3. And the luminosity of the blackbody component, as indicated in the Table 3, varies from $(0.2–3.1) \times 10^{39}$ erg s$^{-1}$, which is notably higher than the Eddington luminosity for a neutron star accretor. This implies an intrinsically high luminosity from the accretion column and a high accretion rate, implying a large thickness of the accretion disc on its inner edge. This results in the reprocessing of a significant part of pulsar radiation in the disc observed as the soft excess. To account for this, either a very strong magnetic field or reprocessing by an optically thick curtain is required (Mushtukov et al. 2015, 2017). However, such extreme luminosities would inevitably result in significant radiation pressure, leading to enormous mass loss. This lost mass would form an optically thick photosphere at a certain distance from the centre, which then reprocesses the emission and contributes to the observed soft excess in the spectrum (Qiu & Feng 2021). The current work can not very firmly establish the origin of the soft excess, but it shows that, empirically, the spectral characteristics of the ULXPs are quite similar to the local X-ray pulsars. The soft component, which is considered to be emitted from the accretion disc, is super Eddington in our model, just as it is in the models employed in several previous works. The spectral model employed in this work is however, simpler with only two components instead of three, one of which is a disc blackbody with an ad-hoc power-law radial dependence of temperature. In these models (Robba et al. 2021; Brightman et al. 2022), the cooler component likely represents the emission from the outflowing winds from the disc due to the mass loss caused by radiation pressure. Same can be true for the blackbody components measured in our model. However, an additional hot blackbody component is not required here.

Iron emission lines observed in X-ray binary spectra typically arise from the interaction between X-rays emitted by the binary system and surrounding material. This surrounding material can be at various locations, including accretion discs (LMXB systems), the high-mass companion star's stellar wind, circumstellar shells situated at a distance from the star, material along the observer's line of sight, and the accretion column (Naik & Paul 2003; Cackett et al. 2010; Fabian & Ross 2010; Paul 2017; Pradhan et al. 2018). For ULXPs, we did not detect a significant iron emission line in the 6–7 keV energy range and have determined the upper limits on the equivalent width for the 6.4 keV line of ∼60–90 eV. Till now there are only two ULX sources (M82 X-1 and NGC 7456 ULX1) in which tentative detection of iron emission line has been reported. In M82 X-1, a weak broadened iron emission line is observed with EW of $59^{+25}_{-27}$ eV (Caballero-García 2011) and in NGC 7456 ULX1, iron emission line is observed with EW of $2000^{+1500}_{-1100}$ eV (Mondal et al. 2021), which is much higher than the M82 X-1. The ULX pulsar like SMC X–3 which stands at the threshold between sub-Eddington X-ray pulsars and ULXPs, has shown a clear detection of iron line features with an equivalent width of ∼70 eV along with low absorption column density of $1-2 \times 10^{21}$ cm$^{-2}$ (Tsygankov et al. 2017; Koliopanos & Vasilopoulos 2018). Galactic ULXP candidate, RX J0209.6−7427 also show a presence of Fe K-alpha emission line (Vasilopoulos et al. 2020). Swift J0243.6+6124, another Galactic ULXP candidate, shows evolution in the iron emission line from a narrow 6.4 keV line at the sub-Eddington regime to a broad line in the super-Eddington regime (Jaisawal et al. 2019; Bykov et al. 2022).

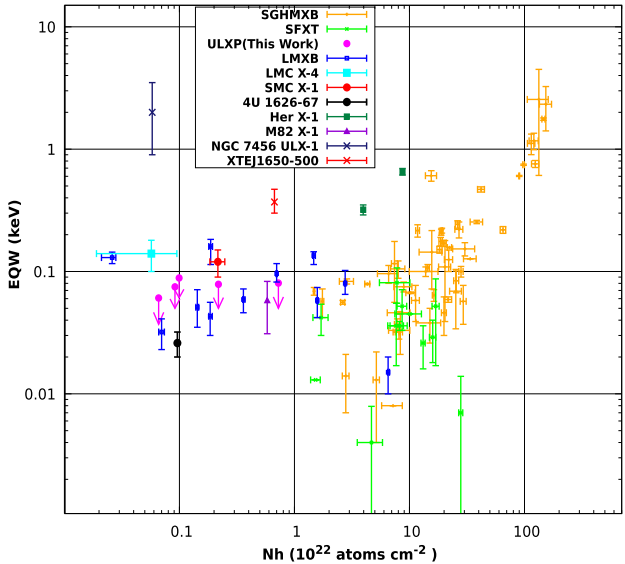

**Figure 2.** Comparative studies of hydrogen column density ($N_\mathrm{H}$: number of hydrogen atoms per squared cm projected along the line of sight) and equivalent width (EQW) of iron K-alpha line for SgHMXBs, SFXTs (Pradhan et al. 2018), LMXBs (Ng, C. et al. 2010), LMC X-4, SMC X-1 (Paul et al. 2002), Her X-1 (Naik & Paul 2003), 4U 1626−67 (Sharma, Jain & Paul 2023b), M82 X-1 (Caballero-García 2011), NGC 7456 ULX1 (Mondal et al. 2021), XTE J1650-500 (Walton et al. 2012), and ULXPs. The SgHMXBs are indicated in orange, SXFTs in green, LMXBs in blue, LMC X-4 in cyan, SMC X-1 in red, Her X-1 in dark green, 4U 1626−67 in black, M82 X-1 in purple, XTE J1650-500 in red (different point type), and NGC 7456 ULX1 in midnight-blue colour. The ULXPs measurements are indicated in magenta points.

In Fig. 2, we show the absorption column density and the equivalent width of iron line observed in ULXPs to those of Supergiant high-mass X-ray binaries (SgHMXBs), supergiant fast X-ray transients (SFXTs; Pradhan et al. 2018), low-mass X-ray binaries (LMXBs; Ng, C. et al. 2010), Galactic stellar mass black hole binary XTE J1650-500 (Walton et al. 2012), and two ULX sources M82 X-1 (Caballero-García 2011) and NGC 7456 ULX1 (Mondal et al. 2021). Despite the ULXPs accreting at super-Eddington rates, they exhibit low absorption column density. At least two of the ULXPs M82 X2 and M51 ULX7 have high-mass companion of mass $M_\mathrm{c} > 5.2\,\mathrm{M}_\odot$ (Bachetti et al. 2014) and $M_\mathrm{c} > 8\,\mathrm{M}_\odot$ (Castillo et al. 2020), respectively and long duration eclipses indicate high mass companion for many, if not all ULXs (Urquhart & Soria 2016). In comparison with Galactic X-ray binaries, the absorption column density in ULXPs is found to be on the lower side of that for the SgHMXBs. Be-X-ray pulsars, however, can have very low column density. On the other hand, upper limits of equivalent width for the iron K-alpha line determined for ULXPs are not in conflict with the low equivalent width generally observed in galactic LMXBs and disc-fed pulsars. In some galactic high-mass X-ray binaries, the iron line has an equivalent width ranging from several hundred eV up to 1.5 keV (Pradhan et al. 2018). Despite having high mass companion and small binary orbit, the reason for very low absorption column density and the absence of iron emission lines in ULXP is still to be understood.

Three ULXPs, M51 ULX7, M82 X-2, and NGC 5907 ULX1 have short orbital periods between 2 and 5 d, suggesting compact orbits. Given that ULXPs have about 2–4 orders of magnitude higher X-ray luminosity than the persistent SgHMXBs, if the accretion were driven by stellar wind, it would require very dense stellar winds, resulting in higher absorption column densities and strong emission lines. However, the lack of strong iron line and low column density indicates the accretion process in ULXPs is likely due to Roche lobe overflow. X-ray pulsars like LMC X-4, SMC X-1, and 4U 1626–67 are believed to have accretion driven by Roche Lobe overflow and show relatively low column density and iron line strength (e.g. Paul et al. 2002; Sharma et al. 2023b).

In the ULX regime, the entire magnetosphere becomes dominated by optically thick material, which reprocesses all the primary photons of the accretion column (Mushtukov et al. 2017). This essentially erases the pulsation information and the spectral characteristics of the accretion column emission. Therefore, it is possible that other ULX sources with no detection of pulsed emission may also harbour neutron star at their central core (King, Lasota & Kluźniak 2017; Pintore et al. 2017 ; Koliopanos & Vasilopoulos 2018; Walton et al. 2018b). This study puts forward a comprehensive comparison between ULXPs and local X-ray pulsars which may be further helpful in understanding other ULX sources.

## ACKNOWLEDGEMENTS

We thank the anonymous referees for their insightful comments and suggestions. This research has made use of data obtained with *NuSTAR*, a project led by Caltech, funded by NASA and managed by NASA/JPL, and has utilized the NUSTARDAS software package, jointly developed by the ASDC (Italy) and Caltech (USA). This work has also made use of data obtained with *XMM–Newton*, an ESA science mission with instruments and contributions directly funded by ESA Member States. This research has made use of archival data and software provided by NASA's High Energy Astrophysics Science Archive Research Center (HEASARC), which is a service of the Astrophysics Science Division at NASA/GSFC.

## DATA AVAILABILITY

Data used in this work can be accessed through the HEASARC archive at https://heasarc.gsfc.nasa.gov/cgi-bin/W3Browse/w3browse.pl.

## APPENDIX A: OBSERVED PROPERTIES OF ULXPS IN PREVIOUS LITERATURE

**Table A1.** Observed properties of ULXPs in previous literature.

| Source | NGC 7793 P13[a] | NGC 5907 ULX1[b] | M51 ULX-7[c] | NGC 300 ULX1[d] | NGC 1313 X-2[e] |
|---|---|---|---|---|---|
| $N_H\|_{GAL}$ ($10^{20}$cm$^{-2}$) | 1.2 [f] | 1.38 [f] | – | 11 ± 1 | – |
| $N_H\|_{ISM}$ ($10^{20}$cm$^{-2}$) | 8 ± 1 | $85^{+14}_{-12}$ | $3.9^{+1.3}_{-0.6}$ | 75 ± 7 | $32^{+3}_{-2}$ |
| CV$_{frac}$ | – | – | – | 0.85 ± 0.03 | – |
| $kT_{diskbb}$ (keV) | $0.45^{+0.03}_{-0.04}$ | $0.30^{+0.09}_{-0.06}$ | $0.33^{+0.03}_{-0.03}$ | $0.178^{+0.008}_{-0.007}$ | 0.21 ± 0.02 |
| $kT_{diskpbb}$ (keV) | 2.0 ± 0.3 | – | $2.26^{+0.54}_{-0.17}$ | – | $1.57^{+0.07}_{-0.05}$ |
| $p$ | > 1.3 | – | $0.99_{-0.25}$[g] | – | 0.57 ± 0.02 |
| $\Gamma$ | $3.7^{+0.9}_{-1.1}$ | $0.83^{+0.13}_{-0.15}$ | 0.8 [f] | 1.52 ± 0.03 | – |
| $E_{fold}$ (keV) | – | $5.3^{+0.7}_{-0.6}$ | – | 7.0 ± 0.3 | – |
| $E_{cut}$ (keV) | – | – | 8.1 [f] | 5.6 ± 0.2 | – |
| References | Walton et al. (2018a) | Fürst et al. (2017) | Brightman et al. (2022) | Carpano et al. (2018) | Robba et al. (2021) |

[a]Model used: tbnew ∗ tbnew ∗ (diskbb + diskpbb ∗ SIMPL).
[b]Model used: tbabs ∗ tbabs ∗ (diskbb + cutoffpl).
[c]Model used: tbabs ∗ (diskbb + diskpbb + cutoffpl).
[d]Model used: tbabs ∗ pcfabs ∗ (diskbb + highecut∗powerlaw).
[e]Model used: tbabs ∗ (diskbb + diskpbb).
[f]These parameters are fixed to corresponding values.
[g]This parameter reached to its upper bound in fit.

This paper has been typeset from a TEX/LATEX file prepared by the author.